\documentclass[manuscript]{aastex}

\usepackage{amssymb}
\usepackage{color}

\begin{document}


\title{Einstein static universe on the brane supported by  
extended Chaplygin gas}


\author{Y. Heydarzade\altaffilmark{1}, ~~F. Darabi\altaffilmark{2} ~~and ~~K. Atazadeh\altaffilmark{3}}
\affil{Department of Physics, Azarbaijan Shahid Madani University , Tabriz, 53714-161 Iran.\\
Research Institute for Astronomy and Astrophysics of Maragha (RIAAM), Maragha 55134-441, Iran.}

\altaffiltext{1}{email: heydarzade@azaruniv.edu}
\altaffiltext{2}{email: f.darabi@azaruniv.edu}
\altaffiltext{3}{email: atazadeh@azaruniv.edu}

%

\begin{abstract}
 We study the cosmological models in which an extended Chaplygin gas universe is merged with the braneworld scenario. In particular, we examine the realization of  Einstein static universe on the brane embedded in a non-constant curvature
bulk space and perform a detailed perturbation analysis. We extract the stability conditions and find their impacts on the geometric equation of state parameter  and the spatial curvature of the universe.
\end{abstract}


\maketitle

\section{Introduction}
According to observations from different areas of cosmology, the universe has entered in a 
phase of accelerated expansion in the recent cosmological past  
\citep{Riess:1998cb,Perlmutter:1998np}. Although the incorporation of a cosmological 
constant is the simplest explanation \citep{Peebles:2002gy}, the possibility that the 
acceleration exhibit dynamical features led to two scenario. 

In the first scenario, one can introduce the concept of dark energy, i.e. change the right hand side of the Einstein field equations. This can be realized by a quintessence field
\citep{Ratra:1987rm,Wetterich:1987fm,Liddle:1998xm,Guo:2006ab,Dutta:2009yb}, a phantom field \citep{Caldwell:1999ew,Caldwell:2003vq,Nojiri:2003vn,Onemli:2004mb,
Saridakis:2008fy}, both fields as a quintum field\citep{Guo:2004fq,Zhao:2006mp,Cai:2009zp}, or more complex scenarios like K-essence \citep{ArmendarizPicon:2000ah}, Hordenski \citep{Horndeski1974}, Galileons \citep{Nicolis:2008in,Deffayet:2009wt,Deffayet:2011gz,Leon:2012mt}, holographic dark energy \citep{Hooft1,Hooft2,Hooft3,Hooft4}, etc (for a review the reader is referred to \citep{Copeland:2006wr}).

In the second scenario, one can introduce additional terms in the gravitational Lagrangian, that is modifying the gravitational theory, and consider the dark energy sector as an effective sector of gravitational origin. 
Specifically one can have the $f(R)$ gravity 
\citep{DeFelice:2010aj,Capozziello:2005ku,
Amendola:2006we,Capozziello:2010ih}, the Gauss-Bonnet gravity
\citep{Wheeler:1985nh,Nojiri:2005jg,DeFelice:2008wz,Rastkar:2011sx}, the Weyl gravity
\citep{Mannheim:1988dj,Flanagan:2006ra}, the Lovelock gravity
\citep{Lovelock:1971yv,Deruelle:1989fj}, the Ho\v{r}ava-Lifshitz gravity  \citep{Horava:2008ih,Calcagni:2009ar,Saridakis:2009bv}, 
the nonlinear massive gravity terms
\citep{deRham:2010kj,Hinterbichler:2011tt,deRham:2014zqa,Leon:2013qh}, the $f(T)$ gravity  
\citep{Bengochea:2008gz,Linder:2010py,Atazadeh:2011a,Paliathanasis:2014iva} etc (for a reviews the reader is referred to \citep{Capozziello:2011et,Nojiri:2006ri}). An interesting class of modified gravity also comes from the braneworld scenario, in which the universe is a brane embedded in a higher dimensional spacetime \citep{Rubakov:1983bb,Randall:1999ee,Darabi:2000yg,Brax:2003fv,Koyama:2007rx,
Maeda:2007cb,Bazeia:2008zx,Setare:2008mb,Lombriser:2009xg}, however the gravitational interaction can act on the whole higher dimensional ``bulk'' space. Hence, the universe evolution is determined by the combination of the matter behavior on the brane, plus the effects if the higher dimensional gravity.

In the majority of cosmological models of the first scenario, the dark energy and dark matter belong to different sectors. However, one can propose scenarios where both these sectors are unified in a unique definition. This is indeed achieved by assuming that there is a unique cosmic fluid with an equation of state parameter of a Chaplygin gas type
\citep{Kamenshchik,Bilic:2001cg,Gorini:2002kf} or its extensions
\citep{Bento:2002ps,HeydariFard:2007uw,HeydariFard:2008tm,Ali:2011sv,Pourhassan:2013sw, Kahya:2014fja,Lu:2014rza,Pourhassan:2014ika}, where at early times of universe
evolution behaves as a pressureless fluid (matter-dark matter era), and at late times behaves like the cosmic fluid which mimics the cosmological constant (dark energy era).

In this work, we aim to study the cosmological models in which an extended Chaplygin gas universe, of the first scenario, is merged with the braneworld universe, of the
second scenario. Moreover, 
motivated by the cosmological emergent universe scenario \citep{emergent1,
emergent2,emergent3},
where the big bang singularity is removed
and the Universe is originated from an Einstein static state,  we study the Einstein static universe and its stability   in such a model.
Similar attempts  have already been done in the context of modified theories of gravity such as
$f(R)$  \citep{f(R)1,f(R)2,f(R)3,f(R)4,f(R)5}, $f(T)$
\citep{f(T)1, f(T)2}, Einstein-Cartan theory \citep{Cartan1,Cartan2}, massive gravity \citep{massive1, massive2, massive3},
loop quantum cosmology \citep{Loop1,Loop2,Loop3}, non-minimal kinetic coupled gravity \citep{ata}, Horava-Lifshitz gravity \citep{horava,
yaghoubhorava}, braneworld scenarios \citep{brane1,brane2,brane3,brane4,yaghoubbrane},
induced matter theory \citep{yaghoub} Lyra geometry \citep{Darabi} and
doubly  general relativity\citep{mohsen}. We follow the approach of  
\citep{yaghoubhorava} and \citep{yaghoubbrane,yaghoub,Darabi,mohsen} and extract the stability  
regions in terms of the geometric linear equation of state parameter.  The plan of this work is as follows: In section 2, we present the geometrical setup of the model. In section 3, we perform a detailed analysis of the cosmological perturbations
and focus on the Einstein static universe and extract the conditions for its stability. In section 4, we study some specific
solutions. Finally, in section 5, we discuss on our results.

\section{General Geometrical Setup of the Model}
The effective Einstein-Hilbert action functional for the $4D$ spacetime 
$(\mathcal{M}_{4},g)$
embedded in a $n$-dimensional ambient  space $(\mathcal{M}_{n},\mathcal{G})$ can be derived
from the action 
\begin{equation}
I_{EH}=\frac{1}{2\kappa_{n}^2}\int d^{n}x\sqrt{-\mathcal{G}}\mathcal{R}+ \int_{\Sigma}
d^{4}x\sqrt{-g}\mathcal{L}_{m},
\end{equation}
where $\kappa_{n}^2$ is the bulk space energy scale and $\mathcal{L}_m$ is
the lagrangian of confined matter fields to the brane. The confinement hypothesis
represents that the matter fields are trapped on the four dimensional brane.
Variation of this action with respect to the ambient metric
$\mathcal{G}_{AB}(A,B=0,...,n-1)$ leads to the following Einstein field equations
for the ambient
space\begin{equation}
G_{AB}=8\pi G_{n}S_{AB,}
\end{equation}
where $G_n$ is the ambient  gravitational constant and $S_{AB}$ is the matter energy-momentum tensor.
Using the confinement hypothesis, we have
\begin{equation}
8\pi G_{n}S_{\mu\nu}=8\pi GT_{\mu\nu},~~S_{\mu a}=0,~~S_{ab}=0,
\end{equation}
where $a,b=4,...,n-1$ labels the number of extra dimensions and $T_{\mu\nu}$ is the 
confined matter 
source on the brane.

For obtaining the effective Einstein field equation induced on the brane,
we consider the following geometrical setup.
 Consider the $4D$  background Lorentzian submanifold $\mathcal{M}_{4}$ 
isometrically embedded
in a $n$ dimensional ambient space $\mathcal{M}_{n}$ by a differential map ${\cal 
Y}^{A}:\mathcal{M}_{4}\longrightarrow\mathcal{M}_{n}$ such that
 \begin{eqnarray}
 \label{21}
{\cal G} _{AB} {\cal Y}^{A}_{,\mu } {\cal Y}^{B}_{,\nu}=
\bar{g}_{\mu \nu}  , \hspace{.5 cm} {\cal G}_{AB}{\cal
Y}^{A}_{,\mu}\bar{\cal N}^{B}_{a} = 0  ,\hspace{.5 cm}  {\cal
G}_{AB}\bar{\cal N}^{A}_{a}\bar{\cal N}^{B}_{b} = {g}_{ab},
\end{eqnarray}
where ${\cal G}_{AB}$ $(\bar{g}_{\mu \nu })$ is the metric of the ambient
(brane) space $\mathcal{M}_{n}(\mathcal{M}_{4})$ in which $\{{\cal Y}^{A}\}$ $(\{x^{\mu 
}\})$ is 
the basis
of the ambient (brane), ${\bar{\cal N}^{A}}_{a}$ are $(n-4)$ normal 
unit vectors orthogonal to the brane and  $g_{ab}=\epsilon\delta_{ab}$
in which $\epsilon=\pm1$ represent the two possible signature of each extra dimension. Perturbation of the background
submanifold $\mathcal{M}_{4}$ in a sufficiently small
neighborhood of the brane along an arbitrary transverse direction $\xi^a$ is
given by the following relation
\begin{eqnarray}\label{22}
{\cal Z}^{A}(x^{\mu},\xi^{a}) = {\cal Y}^{A} + ({\cal
L}_{\xi}{\cal Y})^{A}, \label{eq2}
\end{eqnarray}
where ${\cal L}_{\xi^a}$ is the Lie derivative along $\xi^a$ where
$\xi^a$ 
 with $a=5,...,n$ are small
parameters along ${{\cal N}^{A}}_{a}$ parameterizing the  non-compact
extra dimensions.
 By choosing the extra dimensions $\xi^a$ to be orthogonal to the brane, the
gauge
independency is guarantied
\citep{Nash, Jalal}
and we will have  perturbations of the embedding along the
orthogonal extra directions ${{{\cal N}}}^{A}_{a}$ which leads the local
coordinates of the perturbed brane as
\begin{eqnarray}\label{23}
&&{\cal Z}_{,\mu }^{A}(x^{\nu },\xi^a)={\cal Y}_{,\mu }^{A}(x^{\nu })+
\xi^{a}{{{\cal N}}^{A}}_{a,\mu },\nonumber\\
&& {\cal Z}_{,a }^{A}(x^{\nu },\xi^a)={{\cal N}^{A}}_{a}.
\end{eqnarray}
It is seen from equation (\ref{22}) that since the vectors ${{\cal N}}^{A}$ depend only 
on the 
local coordinates $x^{\mu }$,  ${{\cal N}}^{A}={{\cal N}}^{A}(x^\mu)$,
 they do not
propagate along the extra dimensions of the ambient space and we have\begin{equation}\label{24}
{\cal N}^{A}_{a}={\bar{\cal N}^{A}}_{\,\,\,\,\,a}+\xi^{b}\left[{\bar{\cal 
N}^{A}}_{a},{\bar{\cal N}
^{A}}_{b}\right]={\bar{\cal N}^{A}}_{\,\,\,\,\,a}.
\end{equation}
These  considerations give the embedding equations of the
perturbed geometry
as\begin{eqnarray}\label{25}
{\cal G}_{AB}{\cal Z}_{\,\,\ ,\mu }^{A}{\cal
Z}_{\,\,\ ,\nu }^{B}=g_{\mu \nu },\hspace{0.5cm}{\cal G}_{AB}{\cal
Z}_{\,\,\ ,\mu }^{A}{\cal N}_{\,\,\ a}^{B}=g_{\mu a},\hspace{0.5cm}{\cal
G}_{AB}{\cal N}_{\,\,\
a}^{A}%
{\cal N}_{\,\,\ b}^{B}={g}_{ab}.
\end{eqnarray}
where by setting ${{{\cal N}}^{A}}_{a}={\delta^{A}}_{a}$, the metric of the ambient space 
${\cal G}_{
AB}$ in the vicinity of submanifold $\mathcal{M}_{4}$ and in the Gaussian frame can be
written in the following matrix form 
\begin{eqnarray}\label{26}
{\cal G}_{AB}=\left( \!\!\!%
\begin{array}{cc}
g_{\mu \nu }+A_{\mu c}A_{\,\,\nu }^{c} & A_{\mu a} \\
A_{\nu b} & g_{ab}%
\end{array}%
\!\!\!\right) , 
\end{eqnarray}%
which leads to the  following line element for the ambient space
 \begin{equation}\label{27}
dS^{2}={\cal G}_{AB}d{\cal Z}^{A}d{\cal Z}^{B}=g_{\mu \nu
}(x^{\alpha },\xi^a)dx^{\mu }dx^{\nu }+g_{ab}d\xi^{a}d\xi^{b},  \end{equation}
where
\begin{eqnarray}\label{28}
g_{\mu \nu }=\bar{g}_{\mu \nu }-2\xi ^{a}\bar{K}_{\mu \nu a}+\xi
^{a}\xi ^{b}%
\bar{g}^{\alpha \beta }\bar{K}_{\mu \alpha a}\bar{K}_{\nu \beta
b},
\end{eqnarray}%
is the metric of the perturbed brane, or the first fundamental form, and\begin{eqnarray}\label{29}
\bar{K}_{\mu \nu a}=-{\cal G}_{AB}{\cal Y}_{\,\,\,,\mu }^{A}{\cal
N}_{\,\,\ a;\nu }^{B}=-\frac{1}{2}\frac{\partial
g_{\mu \nu }}{\partial\xi^{a} },
\end{eqnarray}%
is the extrinsic curvature of the original brane, or the second
fundamental form. We use the notation $A_{\mu c}=\xi ^{d}A_{\mu
cd}$  where \begin{equation}\label{210}
A_{\mu cd}={\cal G}_{AB}{\cal N}_{\,\,\ d;\mu }^{A}{\cal N}_{\,\,\
c}^{B}=%
\bar{A}_{\mu cd},  
\end{equation}%
is known as the twisting vector fields,
or the normal fundamental form.  Any fixed $\xi^a$ denotes a new perturbed
brane in which we can define an extrinsic curvature for this perturbed
brane similar to the
original one in the following form
\begin{eqnarray}\label{211}
\tilde{K}_{\mu \nu a}=-{\cal G}_{AB}{\cal Z}_{\,\,\ ,\mu
}^{A}{\cal
N}%
_{\,\,\ a;\nu }^{B}=\bar{K}_{\mu \nu a}-\xi ^{b}\left(
\bar{K}_{\mu
\gamma a}%
\bar{K}_{\,\,\ \nu b}^{\gamma }+A_{\mu ca}A_{\,\,\ b\nu
}^{c}\right).
\end{eqnarray}
Note that the definitions (\ref{26}), (\ref{28}) and (\ref{211}) require
\begin{eqnarray}
\tilde{K}_{\mu \nu a}=-\frac{1}{2}\frac{\partial {\cal G}_{\mu
\nu
}}{%
\partial \xi ^{a}}. 
\end{eqnarray}%
In the presence of gauge fields $A_{\mu a}$,
 the embedded family of submanifolds are tilted with respect to the
normal vector ${\cal N} ^{A}$. According to our geometrical construction, the
original brane is orthogonal to the normal vector ${\cal N}^{A}.$
However,  the equation (\ref{25})  shows that this is not the case for the
deformed geometry. Then, we change the embedding coordinates to the following
form \begin{eqnarray}\label{212}
{\cal X}_{,\mu }^{A}={\cal Z}_{,\mu }^{A}-g^{ab}{\cal
N}_{a}^{A}A_{b\mu }, 
\end{eqnarray}
where the coordinates ${\cal X}^{A}$ describe a new family of embedded
submanifolds whose members are always orthogonal to ${\cal N}^{A}$.
In this coordinates the embedding equations of the perturbed brane
is similar to the original one, represented by the equation (\ref{21}), 
so that the coordinates ${\cal Y}^{A}$ is replaced by ${\cal X}^{A}$. The
embedding of the local coordinates ${\cal X}^{A}$ are suitable than ${\cal
Z}^{A}$ for obtaining
the induced Einstein field equations on the brane. The extrinsic
curvature of a perturbed brane in these coordinates,
 becomes
\begin{eqnarray}\label{213}
K_{\mu \nu a}=-{\cal G}_{AB}{\cal X}_{,\mu }^{A}{\cal N}_{a;\nu
}^{B}=\bar{K}%
_{\mu \nu a}-\xi ^{b}\bar{K}_{\mu \gamma a}\bar{K}_{\,\,\nu
b}^{\gamma
}=-\frac{1}{2}\frac{\partial g_{\mu \nu }}{\partial \xi ^{a}},
\end{eqnarray}
which is the generalized York's relation representing the
propagation of the extrinsic curvature due to the metric propagation in the 
direction of 
extra dimensions. The components of
the Riemann tensor of the ambient space in the embedding vielbein
$\{{\cal X}^{A}_{, \alpha}, {\cal N}^A_a \}$, yield  the
Gauss-Codazzi equations  \citep{Eisenhart}
as\begin{eqnarray}\label{215}
R_{\alpha \beta \gamma \delta}=2g^{ab}K_{\alpha[ \gamma
a}K_{\delta] \beta b}+{\cal R}_{ABCD}{\cal X} ^{A}_{,\alpha}{\cal
X} ^{B}_{,\beta}{\cal X} ^{C}_{,\gamma} {\cal X}^{D}_{,\delta},
\end{eqnarray}
\begin{eqnarray}\label{216}
2K_{\alpha [\gamma c; \delta]}=2g^{ab}A_{[\gamma ac}K_{ \delta]
\alpha b}+{\cal R}_{ABCD}{\cal X} ^{A}_{,\alpha} {\cal N}^{B}_{c}
{\cal X} ^{C}_{,\gamma} {\cal X}^{D}_{,\delta},
\end{eqnarray}
where ${\cal R}_{ABCD}$ and $R_{\alpha\beta\gamma\delta}$ are the
Riemann tensors of the ambient space and the perturbed brane, respectively.
 The Ricci tensor is obtainable by contracting the Gauss equation (\ref{215}) as
\begin{eqnarray}\label{217}
R_{\mu\nu}=(K_{\mu\alpha c}K_{\nu}^{\,\,\,\,\alpha c}-K_{c} K_{\mu
\nu }^{\,\,\,\ c})+{\cal R}_{AB} {\cal X}^{A}_{,\mu} {\cal
X}^{B}_{,\nu}-g^{ab}{\cal R}_{ABCD}{\cal N}^{A}_{a}{\cal
X}^{B}_{,\mu}{\cal X}^{C}_{,\nu}{\cal N}^{D}_{b}.
\end{eqnarray}
The next contraction will give the Ricci scalar
as
\begin{equation} \label{218}
R={\cal R}+(K_{a\mu \nu }K^{a\mu \nu }-K_{a}K^{a})-2g^{ab}{\cal R}_{AB}{\cal 
N}_{a}^{A}{\cal N}_{b}^{B}+g^{ad}g^{bc}{\cal R}_{ABCD}{\cal N}_{a}^{A}{\cal 
N}_{b}^{B}{\cal N}_{c}^{C}{\cal N}_{d}^{D}, 
\end{equation}
where $K_{a}\equiv g^{\mu \nu }K_{a\mu \nu }$.
Then, by using equations
(\ref{217}) and (\ref{218}), we can obtain the following relation
between the Einstein tensors of the ambient space and brane
\begin{equation} \label{219}
G_{AB}{\cal X}_{,\mu }^{A}{\cal X}_{,\nu }^{B}=G_{\mu \nu }-Q_{\mu \nu }-g^{ab}
{\cal R}_{AB}{\cal N}_{a}^{A}{\cal N}_{b}^{B}g_{\mu \nu }+g^{ab}{\cal R}
_{ABCD}{\cal N}_{a}^{A}{\cal X}_{\mu }^{B}{\cal X}_{\nu }^{C}{\cal N}
_{b}^{D}, 
\end{equation}
where $G_{AB}$ and $G_{\mu \nu }$ are the Einstein tensors of the
ambient space and brane respectively, and
the new  quantity $Q_{\mu\nu}$ as\begin{equation}\label{220}
Q_{\mu \nu }=g^{ab}(K_{a\mu }^{\,\,\,\,\,\,\gamma }K_{\gamma \nu
b}-K_{a}K_{\mu \nu b})-\frac{1}{2}(K_{a\mu \nu }K^{a\mu \nu }-K_{a}K^{a})g_{\mu \nu },
\end{equation}
 is an independent conserved
geometrical quantity, i.e. $\nabla _{\mu}Q^{\mu \nu }=0$ \citep{Maia:2004fq}.

Using the decomposition of the Riemann tensor of the ambient space into the
Weyl curvature tensor, the Ricci tensor and the scalar curvature as 
\begin{equation}
\mathcal{R}_{ABCD}=C_{ABCD}-\frac{2}{n-2}\left(\mathcal{G}_{B[D}\mathcal{R}_{C]A}-
\mathcal{G}_{A[D}\mathcal{R}_{C]B}\right)-
\frac{2}{(n-1)(n-2)} \mathcal{R}(\mathcal{G}_{A[D}\mathcal{R}_{C]B}),
\end{equation}
we obtain the  four dimensional induced Einstein equation on the brane as
\begin{eqnarray} 
\label{222}
G_{\mu \nu }&=&G_{AB}{\cal X}_{,\mu }^{A}{\cal X}_{,\nu }^{B}+Q_{\mu \nu 
}-\mathcal{E}_{\mu\nu}+\frac{n-3}{n-2}g^{ab}
{\cal R}_{AB}{\cal N}_{a}^{A}{\cal N}_{b}^{B}g_{\mu \nu}\nonumber\\
&&- \frac{n-4}{n-2}\mathcal{R}_{AB}{\cal X}_{,\mu }^{A}{\cal X}_{,\nu 
}^{B}+\frac{n-4}{(n-1)(n-2)}\mathcal{R}g_{\mu\nu}, 
\end{eqnarray}
where ${\cal E}_{\mu \nu }=g^{ab}{\cal C}_{ABCD}{\cal X}
_{,\mu }^{A}{\cal N}_{a}^{B}{\cal N}_{b}^{C}{\cal X}_{,\nu }^{D}$
is the electric part of
the Weyl tensor of the ambient space ${\cal C}_{ABCD}$. The electric part of the Weyl tensor 
is well 
known from the brane point of view. It represents a traceless matter, denoted by dark 
radiation or 
Weyl matter where for a constant curvature ambient space, we have $\mathcal{E}_{\mu\nu}=0$.

Then, the induced Einstein equation
in a non-constant curvature and Ricci flat ambient space (i.e. $\mathcal{E}_{\mu\nu}\neq0$
and $\mathcal{R}_{AB}=0$)   
will be 
\begin{equation}\label{G}
G_{\mu\nu}=T_{\mu\nu}+Q_{\mu\nu}-\mathcal{E}_{\mu\nu},
\end{equation}
where $T_{\mu\nu}$ is the confined matter source on the brane.

In a cosmological setup,  for the purpose of  embedding of the $FRW$ brane in a five dimensional ambient space, we consider the metric
of
\begin{equation}\label{ds}
 ds^{2}=-dt^{2}+a(t)^{2}(\frac{dr^2}{1-kr^2}+r^{2}d\Omega^2),
 \end{equation}
 where $a(t)$ is the cosmic scale factor and $k=+1, -1$ or $0$ corresponds
 to the closed, open or flat universes.
The confined matter source on the brane $T_{\mu\nu}$ can be considered as a perfect fluid 
given in 
co-moving coordinates
by\begin{equation}\label{T}
 T_{\mu\nu}=(\rho + p)u_{\mu}u_{\nu}+ pg_{\mu\nu},
 \end{equation}
where $u_{\alpha}=\delta^{0}_{\alpha}$, $\rho$ and $p$ are energy density
and isotropic pressure, respectively.
For the confined  extended Chaplygin gas on brane, $p(t)$ has the form of
\begin{equation}\label{pt}
p=\sum_{i=1}^{n} A_{i}\rho^{i}-\frac{B}{{\rho}^\alpha},
\end{equation}
where $A_i$ and $B$ are constants \citep{Kahya:2014fja,Pourhassan:2014ika,Lu:2014rza}.
This model is reduced to generalized Chaplygin gas model introduced in \citep{Kamenshchik}
and elaborated in \citep{Bento:2002ps} with $A_{i}=0$. Also, it is reduced to the original Chaplygin gas scenario with $A_{i}=0$ and $ \alpha=1$.

In order to obtain the components of $Q_{\mu\nu}$, we need to evaluate the
components of the extrinsic curvature $K_{\mu\nu}$.   Using the Codazzi equation, we obtain  
\begin{eqnarray}\label{
K}
 &&K_{00}=-\frac{1}{\dot a}\frac{d}{dt}\left(\frac{b}{a}\right),\nonumber\\
 &&K_{ij}=\frac{b}{a^2}g_{ij},\ i,j=1,2,3.
 \end{eqnarray}
where dot denotes  the derivative with respect to cosmic time $t$ and $b=b(t)$ is an arbitrary 
function \citep{Maia:2004fq, Maia:1988df}.
By defining the new parameters $h:=\frac{\dot b}{b}$ and $H:=\frac{\dot a}{a}$
the components of $Q_{\mu\nu}$ represented by equation (\ref{220}), 
will be
\begin{eqnarray}\label{Q1}
&& Q_{00}=\frac{3b^2}{a^4},\nonumber\\
&&Q_{ij}=-\frac{b^2}{a^4}\left(\frac{2h}{H}-1\right)g_{ij}.
\end{eqnarray}
Similar to the confined source $T_{\mu\nu}$, the geometric energy-momentum tensor 
$Q_{\mu\nu}$ can 
be identified as a perfect fluid\citep{Maia:2004fq}
\begin{equation}\label{Q2}
 Q_{\mu\nu}=(\rho_{g} + p_{g})u_{\mu}u_{\nu}+ p_{g}g_{\mu\nu},
\end{equation}
where the $\rho_{g}$ and  $p_{g}$ denoting the ``geometric energy density"
and ``geometric pressure", respectively (the index $g$ stands for ``geometric").
Then, using the equations  (\ref{Q1}) and (\ref{Q2}), we will have
 \begin{eqnarray}\label{rho}
 &&\rho_{g}=\frac{3b^2}{a^4},\nonumber\\
 &&p_{g}=-\frac{b^2}{a^4}\left(\frac{2h}{H}-1\right).
 \end{eqnarray}
Also, we consider the geometric fluid to have a barotropic equation of state 
$p_{g}=\omega_{
g}\rho_{g}$ where  $\omega_{g}$ is the geometric
equation of state parameter and generally can be a function of time. Using
 equations (\ref{rho}) and the equation of state of the geometric fluid,
we obtain the following equation for $b(t)$  in terms of the scale factor
$a(t)$ and the equation of state parameter $\omega_g$ as
 \begin{equation}\label{dotb}
\frac{\dot b}{b}=\frac{1}{2}\left(1-3\omega_{g}\right)\frac{\dot a}{a},
\end{equation}
which cannot easily be solved because $\omega_{g}$ is not known.
However, in the case of studying the Einstein static universe, a simple and useful consideration 
can be $\omega_{g}=\omega_{0g}=constant$ leading to 
a general solution for the equation (\ref{dotb})
as
\begin{equation}\label{b}
b=b_{0}\left(\frac{a}{a_0}\right)^{\frac{1}{2}(1-3\omega_{0g})},
\end{equation}
where $a_0=constant$ is the scale factor of Einstein static universe and $b_0$ is an 
integration 
constant representing the curvature warp of this
universe. Substituting  equation (\ref{b}) into  equations (\ref{Q1})
gives the geometric fluid component in terms of $b_0$, $a_0$ and $a(t)$ as
\begin{eqnarray}\label{211b}
&&Q_{00}(t)=\frac{3b_{0}^{2}}{a_{0}^{1-3\omega_{g}}}a^{-3(1+\omega_{g})},\nonumber\\
&&Q_{ij}(t)=3\omega_{g}\frac{b_{0}^{2}}{a_{0}^{1-3\omega_{g}}}a^{-3(1+\omega_{g})}g_{ij},
\end{eqnarray}
and consequently using equations (\ref{rho}), 
we get
\begin{eqnarray}\label{rhot}
&&\rho_{g}(t)=\frac{3b_{0}^{2}}{a_{0}^{1-\omega_{g}}}a^{-3(1+\omega_{g})},\nonumber\\
&&p_{g}(t)=3\omega_{g}\frac{b_{0}^{2}}{a_{0}^{1-3\omega_{g}}}a^{-3(1+\omega_{g})}.
\end{eqnarray}
For the Einstein static universe, $a=a_{0}=constant$, the  geometric
fluid components are
as follows\begin{eqnarray}\label{Q3}
&&Q_{00}(a_0)=\frac{3b_{0}^2}{a_{0}^4},\nonumber\\
&&Q_{ij}(a_0)=3\omega_{g}\frac{b_{0}^2}{a_{0}^4}g_{ij}.
\end{eqnarray}
Consequently, using equations (\ref{Q3}), the  geometric energy density and isotropic 
pressure will be\begin{eqnarray}\label{rho0}
&&\rho_{0g}=\rho_{g}(a_0)=\frac{3b_{0}^2}{a_{0}^4},\nonumber\\
&&p_{0g} =p_{g}(a_{0})=\frac{3\omega_g b_{0}^2}{a_{0}^4}.
\end{eqnarray}
Using equations (\ref{rhot}) and (\ref{Q2}), the induced Einstein equation on the brane 
(\ref{G}) 
give us the following equation for the confined energy
density
\begin{equation}\label{rhot1}
\rho(t)=3\left(\frac{\dot 
a}{a}\right)^{2}+\frac{3k}{a^2}-\frac{3b_{0}^2}{a_{0}^{1-3\omega_{g}}}a^{-
3(1+\omega_{g})}+\frac{\mu}{a^4},
\end{equation}
where $\mu$ is an integration constant which mathematically can be positive or negative depending on the geometry of the bulk \citep{Mukohyama1999bb}. The standard big-bang cosmology does not include the
third and fourth terms in  the right hand of equation (\ref{rhot1}). The third term comes
from the extrinsic geometry of the embedded 
 brane through the quantity $Q_{\mu\nu}$.
The fourth term  which scales just like as the radiation with
a constant $\mu$, is known as the dark radiation arising from  the electric part of
the Weyl tensor of the ambient space $\mathcal{E}_{\mu\nu}$. 
Both positive and negative valuesµ for $\mu$ are possible mathematically. On the other hand, dark radiation has influence on  both of the big-bang
nucleosynthesis and the cosmic microwave background. Then, one can determine  both the magnitude and sign of the dark radiation using the constraints
coming from the observations related to the big-bang
nucleosynthesis and the cosmic microwave background \citep{DR:2002,
DR1:2001hh}. For the Einstein static universe, the  equation (\ref{rhot1}) takes the following form
\begin{equation}\label{rho01}
\rho_{0}=\rho(a_0)=\frac{3k}{a_{0}^2}-\frac{3b_{0}^2}{a_{0}^4}+\frac{\mu}{a_{0}^4}.
\end{equation}

Similarly, the confined isotropic pressure component can be obtained from  equations 
(\ref{G}), (\ref{Q2}) and (\ref{rhot}) as
\begin{equation}\label{pt1}
p(t)=-2\frac{\ddot a}{a}-\left(\frac{\dot 
a}{a}\right)^{2}-\frac{k}{a^2}-\frac{3b_{0}^{2}\omega_{g}}
{a_{0}^{1-3\omega_{g}}}a^{-3\left(1+\omega_{g}\right)}+\frac{\mu}{a^4},
\end{equation}
leading to
\begin{equation}\label{p0}
\sum_{i=1}^{n} 
A_{i}\rho_{0}^{i}-\frac{B}{{\rho_{0}}^\alpha}=-\frac{k}{a_{0}^2}-\frac{3b_{0}^{2}\omega_{g
}}{a_{0}^4}+\frac{\mu}{a_{0}^4},
\end{equation}
for the Einstein static universe.
\section{Perturbations and stability analysis of the Einstein static state}
In what follows, we consider  linear homogeneous scalar perturbations around
the Einstein static universe, given in equations (\ref{rho01}) and (\ref{p0}), and
explore its stability against these perturbations. Thus, the  perturbation in the cosmic 
scale 
factor $a(t)$ and the confined energy density $\rho(t)$ depend only on
time can be represented by
\begin{eqnarray}\label{a}
&&a(t)\rightarrow a_{0}(1+\delta a(t)),\nonumber\\
&&\rho(t)\rightarrow \rho_{0}(1+\delta \rho(t)).
\end{eqnarray}
Substituting these equations in the equation (\ref{rhot1}) with subtracting $\rho_0$
and linearizing the result
give the following equation
\begin{equation}\label{deltarho}
\rho_{0}\delta 
\rho(t)=\left(-\frac{6k}{a_{0}^2}+\frac{9b_{0}^{2}(1+\omega_{g})-4\mu}{a_{0}^{4}}\right)\delta a(t).
\end{equation}
Similarly, one can consider a linear
equation of state $p(t)=\omega\rho(t)$ for confined source. Applying the
same method in obtaining the equation (\ref{deltarho}), on the
equations (\ref{pt1}) 
and (\ref{p0}) we also get
\begin{equation}\label{sigma}
\left(\sum_{i=1}^{n} iA_{i}{\rho_{0}}^{i-1}+\frac{\alpha 
B}{{\rho_{0}}^{\alpha-1}}\right)\rho_{0}\delta\rho=-2\delta \ddot a 
+\left(\frac{2k}{a_{0}^2}+
\frac{9b_{0}^{2}\omega_{g}(1+\omega_{g})-4\mu}{a_{0}^{4}}\right)\delta
a.
\end{equation}
Substituting equation (\ref{deltarho}) in (\ref{sigma}) gives the equation
\begin{eqnarray}\label{ddota}
\ddot{\delta a}+
\frac{1}{2}[\left(-\frac{6k}{a_{0}^2}+
 \frac{9b_{0}^{2}(1+\omega_{g})-4\mu}{a_{0}^{4}}\right) 
 \left(\sum_{i=1}^{n} iA_{i}{\rho_{0}}^{i-1}
 +\frac{\alpha B}{{\rho_{0}}^{\alpha-1}}\right)\nonumber\\
-\left(\frac{2k}{a_{0}^2}+
\frac{9b_{0}^{2}\omega_{g}(1+\omega_{g})-4\mu}{a_{0}^{4}}\right) ]\delta a=0.
\end{eqnarray}
This equation has the solution
\begin{equation}\label{323}
\delta a=C_{1}e^{i\gamma t}+C_{2}e^{-i\gamma t},
\end{equation}
where $C_1$ and $C_2$ are integration constants and $A$ is given by
\begin{equation}\label{324}
\gamma^{2}=\frac{1}{2}\left[\left(-\frac{6k}{a_{0}^2}+
 \frac{9b_{0}^{2}(1+\omega_{g})-4\mu}{a_{0}^{4}}\right) 
 \left(\sum_{i=1}^{n} iA_{i}{\rho_{0}}^{i-1}+\frac{\alpha 
B}{{\rho_{0}}^{\alpha-1}}\right)\nonumber\\
-\left(\frac{2k}{a_{0}^2}+
\frac{9b_{0}^{2}\omega_{g}(1+\omega_{g})-4\mu}{a_{0}^{4}}\right) \right].
\end{equation}
Then, for having oscillating perturbation modes representing the existence of a stable 
Einstein 
static universe, the following condition should be satisfied
\begin{equation}\label{325}
\left(-\frac{6k}{a_{0}^2}+
 \frac{9b_{0}^{2}(1+\omega_{g})-4\mu}{a_{0}^{4}}\right) 
 \left(\sum_{i=1}^{n}iA_{i}{\rho_{0}}^{i-1}+\frac{\alpha 
B}{{\rho_{0}}^{\alpha-1}}\right)\nonumber\\
-\left(\frac{2k}{a_{0}^2}+
\frac{9b_{0}^{2}\omega_{g}(1+\omega_{g})-4\mu}{a_{0}^{4}}\right)>0,
\end{equation}
which can be rewritten as the following form for the geometric equation of state parameter 
 
\begin{equation}\label{general}
\omega_{g}^{2}+\mathcal{L}_{1}\omega_{g}+\mathcal{L}_{2}
<0,
\end{equation}
where  
\begin{eqnarray}
&&\mathcal{L}_{1}=1-\sum_{i=1}^{n}iA_{i}{\rho_{0}}^{i-1}
-\frac{\alpha B}{{\rho_{0}}^{\alpha-1}},\nonumber\\
&&\mathcal{L}_{2}=\frac{2ka_{0}^2 -4\mu}{9b_{0}^{2}}-\left(\sum_{i=1}^{n}iA_{i}{\rho_{0}}^{i-1}
+\frac{\alpha B}{{\rho_{0}}^{\alpha-1}}\right)\left(1- \frac{6ka_{0}^2 +4\mu}{9b_{0}^{2}} 
\right)
\end{eqnarray}
The inequality (\ref{general}) leads to the following acceptable range
\begin{equation}\label{range}
\omega_{g}^{(1)}<\omega_{g}<\omega_{g}^{(2)},
\end{equation}
where 
\begin{eqnarray}\label{328a}
&&\omega_{g}^{(1)}=-\frac{1}{2}\mathcal{L}_{1}-\frac{}{}\frac{1}{2}\sqrt{
\mathcal{L}_{1}^{2}-4\mathcal{L}_{2}},\nonumber\\
&&\omega_{g}^{(2)}=-\frac{1}{2}\mathcal{L}_{1}+\frac{}{}\frac{1}{2}\sqrt{
\mathcal{L}_{1}^{2}-4\mathcal{L}_{2}}.
\end{eqnarray}
It is seen that the following condition also should be satisfied
\begin{equation}\label{general2}
\mathcal{L}_{1}^{2}-4\mathcal{L}_{2}\geq0,
\end{equation}
which results in the class of solutions to be discussed in the following
section.
\section{Some specific Solutions}
\subsection{The case of $B=0$ with $A_{i\geq2}=0$}
By defining $A_1=\omega$, this case reduces to the barotropic equation
of state with 
\begin{eqnarray}\label{328b}
&&\omega_{g}^{(1)}=-\frac{1}{2}+\frac{1}{2}\omega-\frac{1}{2}\sqrt{
(1+\omega)^{2}-\frac{8k(1+3\omega)a_{0}^2 -16\omega\mu}{9b_{0}^2}},\nonumber\\
&&\omega_{g}^{(2)}=-\frac{1}{2}+\frac{1}{2}\omega+\frac{1}{2}\sqrt{
(1+\omega)^{2}-\frac{8k(1+3\omega)a_{0}^2 -16\omega\mu}{9b_{0}^2}},
\end{eqnarray}
where we can investigate this solution with more details by dividing in the following classes. In this case, we recover the results in \citep{yaghoubbrane} by considering a constant curvature bulk, i.e setting $\mu=0$.
\subsubsection{Vacuum dominated state $\omega=-1$.}

The corresponding acceptable range for $\omega_{g}$ is
\begin{equation}\label{330}
-1-\frac{2}{3}\sqrt{\frac{ka_{0}^2 -\mu}{b_{0}^2}}<\omega_{g}<-1+\frac{2}{3}\sqrt{\frac{ka_{0}
^2 -\mu}{b_{0}
^2}},
\end{equation}
where we also should have
\begin{equation}\label{3mmm}
ka_{0}^2 -\mu\geq0.
\end{equation}
Regarding the relation (\ref{3mmm}), it is seen that  for this state to be stable, other than the  range (\ref{330})
on $\omega_{g}$, the  universe
should be flat or positively curved $(k\geq 0)$ for a constant curvature
bulk, i.e for $\mu=0$. In the case of the existence of a non-constant curvature bulk, for $\mu>0$ we require $k>0$ while for $\mu<0$, the zero,
positive and negative values for $k$ are allowed by setting the appropriate
numerical
values for the parameters $k, a_0$ and $\mu$. Then, for a flat universe,  as confirmed by the current observations, a bulk space with negative curvature
parameter $\mu$ is needed.
Moreover, regarding the values of 
parameters $k, a_0, b_0$ and $\mu$, for the case of $ka_{0}^2 -\mu<b_{0}^2$ , we have the total range $-\frac{5}{3}<\omega_{g}<-\frac{1}{3}$ which represents
that the geometric equation of state parameter completely
lies in the strong energy condition (SEC) violating range while for the case
of $ka_{0}^2 -\mu>b_{0}^2$ it may include a range of normal matter respecting the SEC.
\subsubsection{ Radiation dominated state $\omega=\frac{1}{3}$.}

For this case, the acceptable range
is\begin{equation}\label{332}
-\frac{1}{3}-\frac{2}{3}\sqrt{1-\frac{ka_{0}^2 -\frac{1}{3}\mu}{b_{0}^2}}<\omega_{g}<
-\frac{1}{3}+\frac{2}{3}\sqrt{1-\frac{ka_{0}^2 -\frac{1}{3}\mu}{b_{0}^2}}.
\end{equation}
It is also seen that we should have
\begin{equation}\label{333}
1-\frac{ka_{0}^2 -\frac{1}{3}\mu}{b_{0}^2}\geq0,
\end{equation}
which gives  a restriction on the scale factor of Einstein static universe $a_0$, the 
brane curvature warp $b_0$ and bulk curvature parameter $\mu$. Regarding the constraint (\ref{333}), and in a constant curvature bulk $(\mu=0)$, for the case of flat and positive curvature universe $(k\geq0)$, we have the
total range $-1<\omega_{g}<\frac{1}{3}$ representing that the geometrical
equation of state parameter includes both of the normal and
SEC violating range. For a non-constant curvature bulk, deducing such a total
range
for $\omega_g$ requires the numerical values of $k, a_0, b_0$ and $\mu$.
However, for a spatially flat universe, the bulk space can take both of the positive and
negative curvature parameter $\mu$ regarding  the value of
curvature warp of the brane $b_0$ satisfying relation (\ref{333}).
\subsubsection{ Matter dominated state, $\omega=0$}
The acceptable region for $\omega_{g}$ is
\begin{equation}\label{335b}
-\frac{1}{2}-\frac{1}{2}\sqrt{1-\frac{8ka_{0}^2}{9b_{0}^2}}<\omega_{g}<\\
-\frac{1}{2}+\frac{1}{2}\sqrt{1-\frac{8ka_{0}^2}{9b_{0}^2}}, 
\end{equation}
where there is also  an additional condition on the scale factor of Einstein static universe $a_0$ and the curvature warp $b_0$  as 
\begin{equation}\label{336b}
1-\frac{8ka_{0}^2}{9b_{0}^2}\geq0,
\end{equation}
implying that the universe can be flat, positively or negatively curved depending on the values of the scale factor of Einstein static universe $a_0$ and the curvature warp $b_0$.  For the flat and positively curved universe $(k\geq0)$,
regarding the constraint (\ref{336b}) and depending on the values of $k, a_0$
and $b_0$, the total range for the geometric
equation of state parameter will be $-1<\omega_{g}<0$ representing that the geometrical
equation of state parameter includes both of the normal and
SEC violating range. For the case of $k<0$, deducing such a total range for
$\omega_g$ requires the numerical values of the parameters $k, a_0$ and $b_0$
but a larger range including both of the 
SEC respecting and violating range is obtainable.  
\subsection{ The case of $A_{i}=B=0$.} 
The acceptable region for $\omega_{g}$ is
\begin{equation}\label{335}
-\frac{1}{2}-\frac{1}{2}\sqrt{1-\frac{8ka_{0}^2 -16\mu}{9b_{0}^2}}<\omega_{g}<\\
-\frac{1}{2}+\frac{1}{2}\sqrt{1-\frac{8ka_{0}^2 -16\mu}{9b_{0}^2}}, 
\end{equation}
where there is also  an additional condition on the scale factor of Einstein static universe $a_0$, the curvature warp $b_0$ and bulk curvature parameter
$\mu$ as 
\begin{equation}\label{336}
1-\frac{8ka_{0}^2 -16\mu}{9b_{0}^2}\geq0.
\end{equation}
For $\mu=0$, this case corresponds to  a universe supported just by dust matter on the
brane with a geometrical induced matter. Using relations (\ref{general})-(\ref{general2}), this case has the acceptable region as the same as the above 
 matter dominated state, $\omega=0$.
Then, for this case, the stable ESU demands
an equation of state parameter as the equations  (\ref{335b}) and (\ref{336b}).
For a non-constant curvature bulk with $\mu<0$ and $k>0$, we obtain the total range as $-1<\omega_{g}<0$
including both of the 
SEC respecting and violating range. For $\mu>0$ deducing a range
for $\omega_g$ requires the numerical values of $k, a_0, b_0$ and $\mu$.

\subsection{The case of $A_{i}=0$.} 
This case is known as the generalized Chaplygin gas model introduced in \cite{Kamenshchik}
and elaborated in \citep{Bento:2002ps}. For this case, we obtain
\begin{eqnarray}
\omega_{g}^{(1)}=-\frac{1}{2}+\frac{\alpha B }{2\rho_{0}^{\alpha-1}}-\frac{1}{2}
\sqrt{\left(1-\frac{\alpha B}{\rho_{0}^{\alpha-1}}\right)^{2}
-4\left[\frac{2ka_{0}^2 -4\mu}{9b_{0}^{2}}- \frac{\alpha B}{\rho_{0}^{\alpha-1}}\left(
1-\frac{6ka_{0}^2+4\mu}{9b_{0}^{2}}\right)\right]},\nonumber\\
\omega_{g}^{(2)}=-\frac{1}{2}+\frac{\alpha B }{2\rho_{0}^{\alpha-1}}+\frac{1}{2}
\sqrt{\left(1-\frac{\alpha B}{\rho_{0}^{\alpha-1}}\right)^{2}
-4\left[\frac{2ka_{0}^2 -4\mu}{9b_{0}^{2}}- \frac{\alpha B}{\rho_{0}^{\alpha-1}}\left(
1-\frac{6ka_{0}^2 +4\mu}{9b_{0}^{2}}\right)\right]},
\end{eqnarray}
where we also should have
\begin{equation}
\alpha^2 +2\frac{\rho_{0}^{\alpha-1}}{ B }\left[1-\frac{12ka_{0}^{2} 
+8\mu}{9b_{0}^{2}}\right]\alpha+\
\frac{\rho_{0}^{2(\alpha-1)}}{ B^{2} }\left(1- \frac{ 8ka_{0}^2 -16\mu}{9b_{0}^2} \right)\geq 0
\end{equation}
which confines the acceptable range for $\alpha$ as
\begin{equation}\label{327}
\alpha \leq\alpha^{(1)} ~~~\&~~~\alpha\geq\alpha^{(2)},
\end{equation}
where 
\begin{eqnarray}\label{328}
&&\alpha^{(1)}=-\frac{1}{2}\mathcal{M}_{1}-\frac{1}{2}\sqrt{
\mathcal{M}_{1}^{2}-4\mathcal{M}_{2}},\nonumber\\
&&\alpha^{(2)}=-\frac{1}{2}\mathcal{M}_{1}+\frac{1}{2}\sqrt{
\mathcal{M}_{1}^{2}-4\mathcal{M}_{2}},
\end{eqnarray}
and 
\begin{eqnarray}
&&\mathcal{M}_{1}=2\frac{\rho_{0}^{\alpha-1}}{ B }\left[1-\frac{12ka_{0}^{2} +8\mu}{9b_{0}^{2}}\right],\nonumber\\
&&\mathcal{M}_{2}=\frac{\rho_{0}^{2(\alpha-1)}}{ B^{2} }\left(1- \frac{ 8ka_{0}^2 -16\mu}{9b_{0}^2} \right).
\end{eqnarray}
Generally, deducing a
total acceptable range or a specific value for $\omega_g$ requires the values
of parameters $k, a_0, b_0, \rho_0, \alpha,  \mu$ and $B$. However, it is interesting that for the case of constant curvature bulk and flat universe, i.e $k=0=\mu$, we have
\begin{eqnarray}
&&\omega_{g}^{(1)}=-1,\nonumber\\
&&\omega_{g}^{(2)}=\frac{\alpha B }{\rho_{0}^{\alpha-1}},
\end{eqnarray}
 which using the relation (\ref{range}) gives a restriction on $\alpha$ and $B$ values of the model as 
$\alpha B>- \rho_{0}^{\alpha-1}$. 
For this case we also find
\begin{equation}\label{}
\alpha^{(1)}=\alpha^{(2)}=-\frac{\rho_{0}^{\alpha-1}}{B}.
\end{equation}
Consequently, because of weak energy condition which imposes a positive energy
density,  the $\alpha$ parameter can take any positive or negative values with respect to the $B$ values.
Interestingly, for the  case $\alpha^{}=-\frac{\rho_{0}^{\alpha-1}}{B}$,
we have $\omega_{g}=\omega_{g}^{(1)}=\omega_{g}^{(2)}=-1$ which denotes that the geometric
fluid behaves as the cosmological constant. 
\subsection{The case of $A_{i}=0$ with $\alpha=1.$}
This case represents the standard Chaplygin gas model on the brane. For this case we have
\begin{eqnarray}
&&\omega_{g}^{(1)}=-\frac{1}{2}+\frac{B}{2}-\frac{1}{2}
\sqrt{\left(1-B\right)^{2}
-4\left[\frac{2ka_{0}^2 -4\mu}{9b_{0}^{2}}- B\left(1-\frac{6ka_{0}^2 +4\mu}{9b_{0}^{2}}\right)\right]},\nonumber\\
&&\omega_{g}^{(2)}=-\frac{1}{2}+\frac{B}{2}+\frac{1}{2}
\sqrt{\left(1-B\right)^{2}
-4\left[\frac{2ka_{0}^2 -4\mu}{9b_{0}^{2}}- B\left(1-\frac{6ka_{0}^2 +4\mu}{9b_{0}^{2}}\right)\right]},
\end{eqnarray}
which has  interesting solutions for $B=1$ and $B=\frac{1}{3}$. For $B=1$,
we have 
\begin{eqnarray}
&&\omega_{g}^{(1)}=-
\sqrt{1-\frac{8ka_{0}^2}{9b_{0}^{2}}},\nonumber\\
&&\omega_{g}^{(2)}=
\sqrt{1-\frac{8ka_{0}^2}{9b_{0}^{2}}},
\end{eqnarray}
with the additional condition as $1-\frac{8ka_{0}^2}{9b_{0}^2}>0$.
It is interesting
that  the effect of bulk curvature does not appear in this case. For the spatially flat  and positively curved universe, we obtain the total range for $\omega_g$ as $-1<\omega_{g}<1$ including both of the 
SEC respecting and violating range. Similarly, for the case of $B=\frac{1}{3}$, we obtain
\begin{eqnarray}
&&\omega_{g}^{(1)}=-\frac{1}{3}-\frac{2}{3}
\sqrt{1-\frac{3ka_{0}^2 -2\mu}{3b_{0}^{2}}},\nonumber\\
&&\omega_{g}^{(2)}=-\frac{1}{3}+\frac{2}{3}
\sqrt{1-\frac{3ka_{0}^2-2\mu }{3b_{0}^{2}}},
\end{eqnarray}
which requires $1-\frac{3ka_{0}^2 -2\mu}{3b_{0}^2}\geq0$. For a constant curvature bulk  with spatially flat and positively curved brane
universe, i.e for $\mu=0$ with $k\geq0 $,  we obtain the total range as $-1<\omega_{g}<\frac{1}{3}$
including both of the 
SEC respecting and violating range. For a non-constant curvature bulk with $\mu<0$ and $k>0$, we also obtain the total range as $-1<\omega_{g}<\frac{1}{3}$
including both of the 
SEC respecting and violating range. For $\mu>0$ deducing a range
for $\omega_g$ requires the numerical values of $k, a_0, b_0$ and $\mu$.
\section{Final Remarks}

A cosmological model in which an extended Chaplygin gas universe, with the
equation
of state $p=\sum_{i=1}^{n} A_{i}\rho^{i}-\frac{B}{{\rho}^\alpha}$, is merged
with the braneworld scenario has been investigated in this work. A general stability
condition for the Einstein static state on the brane embedded in a general
non-constant curvature bulk space  has been obtained. Moreover, subsets of generalized and standard Chaplygin gas model,   brane with dust matter and with barotropic equation of state solutions have been separately discussed. For each case, we have obtained the stability conditions and their impacts on the geometric equation of state parameter $\omega_g$ as well as the spatial curvature $k$ of the universe
 in terms of the scale factor of Einstein static universe $a_0$, the brane
curvature warp $b_0$ factor and bulk space curvature parameter $\mu$. In
the following, some of our results are represented.

For the case of $B=0$ with $A_{i\geq2}=0$ and defining $A_1 =\omega,$ we are reduced to the barotropic equation
of state. Then, we analyze
the vacuum dominated state ($\omega=-1$), the radiation dominated state 
($\omega=\frac{1}{3}$)
and matter dominated state ($\omega=0$) in both of the constant and non-constant
curvature bulk space. For the case of the vacuum dominated state with constant
curvature bulk ($\mu=0$),
the  universe
should be flat or positively curved $(k\geq 0)$. In the case of the non-constant
curvature bulk, for $\mu>0$ we require $k>0$ while for $\mu<0$, the zero,
positive and negative values for $k$ are allowed. Moreover, respecting to the values of 
parameters $k, a_0, b_0$ and $\mu$, for the case of $ka_{0}^2 -\mu<b_{0}^2$ , we have the total range $-\frac{5}{3}<\omega_{g}<-\frac{1}{3}$ which represents
that the geometric equation of state parameter completely
lies in the strong energy condition (SEC) violating range while for the case
of $ka_{0}^2 -\mu>b_{0}^2$ it may include a range of normal matter respecting the SEC. For the radiation dominated state with $\mu=0$ and $k\geq0$, we have the
total range $-1<\omega_{g}<\frac{1}{3}$ representing that the geometrical
equation of state parameter  includes both of the normal and
SEC violating range. In the case of matter dominated state with $k\geq0$,
the total range is obtained as $-1<\omega_{g}<0$ representing that the geometrical
equation of state parameter includes both of the normal and
SEC violating range.  

For the case of $A_{i}=B=0$ 
with $\mu=0$, the acceptable region for $\omega_g$ is the same as the  
 matter dominated state $\omega=0$ while for a non-constant curvature bulk with $\mu<0$ and $k>0$,  the total range is obtained as $-1<\omega_{g}<0$
including both of the 
SEC respecting and violating range. 

For the case of $A_{i}=0$ which is known as the generalized Chaplygin gas model, a total range for $\omega_g$ is obtained. Specifically, it is shown that for the case of constant curvature bulk and flat universe ($k=0=\mu$), the geometric
fluid behaves as the cosmological constant, i.e $\omega_g=-1$.

For the case of $A_{i}=0$ with $\alpha=1$, representing the standard Chaplygin gas model on the brane, a total range for $\omega_g$ is also obtained. Interestingly, it is shown that by setting
$B=1$, the effect of bulk curvature does not appear in the acceptable range of $\omega_g$.
For $k\geq 0$, the total range  as $-1<\omega_{g}<1$ including both of the 
SEC respecting and violating range is obtained.
Also, by setting $B=\frac{1}{3}$, for $\mu=0$ with $k\geq0 $,  the total range as $-1<\omega_{g}<\frac{1}{3}$
including both of the 
SEC respecting and violating range is obtained. Finally, for a non-constant curvature bulk with $\mu<0$ and $k>0$, we also obtain the total range as $-1<\omega_{g}<\frac{1}{3}$
including both of the 
SEC respecting and violating range.

\section*{Acknowledgment}
This work has been supported financially by Research Institute for Astronomy and Astrophysics of Maragha
(RIAAM) under research project NO.1/4165-92.

\end{document}